\documentclass{aip-cp}

\usepackage[numbers]{natbib}
\usepackage{rotating}
\usepackage{graphicx}
\usepackage{wrapfig}
\usepackage{caption}
\usepackage{url}\usepackage[english]{babel}
\usepackage{etoolbox}

\newcommand{\eVdist}{\kern-0.045em}
\newcommand{\Hair}{\kern0.166667em}
\newcommand{\hair}{\kern0.066667em}

\newcommand{\kev}{\ensuremath{\mbox{ke}\eVdist\mbox{V}}}
\newcommand{\mev}{\ensuremath{\mbox{Me}\eVdist\mbox{V}}}
\newcommand{\gev}{\ensuremath{\mbox{Ge}\eVdist\mbox{V}}}
\newcommand{\tev}{\ensuremath{\mbox{Te}\eVdist\mbox{V}}}
\newcommand{\capitalhyphen}{\raisebox{0.24ex}{\resizebox{0.4em}{\height}{-}}\kern-0.07em}
\newcommand{\hessj}{HESS\,J0632$+$057}

\makeatletter
\patchcmd{\@afterheading}
    {\clubpenalty \@M}{\clubpenalties 3 \@M \@M 0}{}{}
\makeatother

\begin{document}

\title{A Decade of \tev\ Observations of the Gamma-ray Binary \hessj\ with VERITAS}

\author[aff1]{S.\,Schlenstedt\corref{cor1}}
\author{the VERITAS Collaboration}
\eaddress[url]{veritas.sao.arizona.edu}
\corresp[cor1]{Corresponding author: stefan.schlenstedt@desy.de}
\affil[aff1]{DESY, Platanenallee 6, 15738 Zeuthen, Germany}

\maketitle

\begin{abstract}
The gamma-ray binary \hessj\,(VER\,J0633$+$057) has been observed at very-high energies for a decade by all major systems of imaging atmospheric Cherenkov telescopes. We present here new observations taken by the VERITAS observatory during the season 2015-2016. The observations  cover now all phases of the binary orbit (with its period of about 315 days), showing clearly enhancements around phases 0.35 and 0.75. The results are discussed along with simultaneous observations with {\em Swift's} X-Ray Telescope. 
\end{abstract}

\begin{figure}[th]
\centerline{\includegraphics[trim={0 1.8ex 0 0},clip,width=0.95\textwidth]{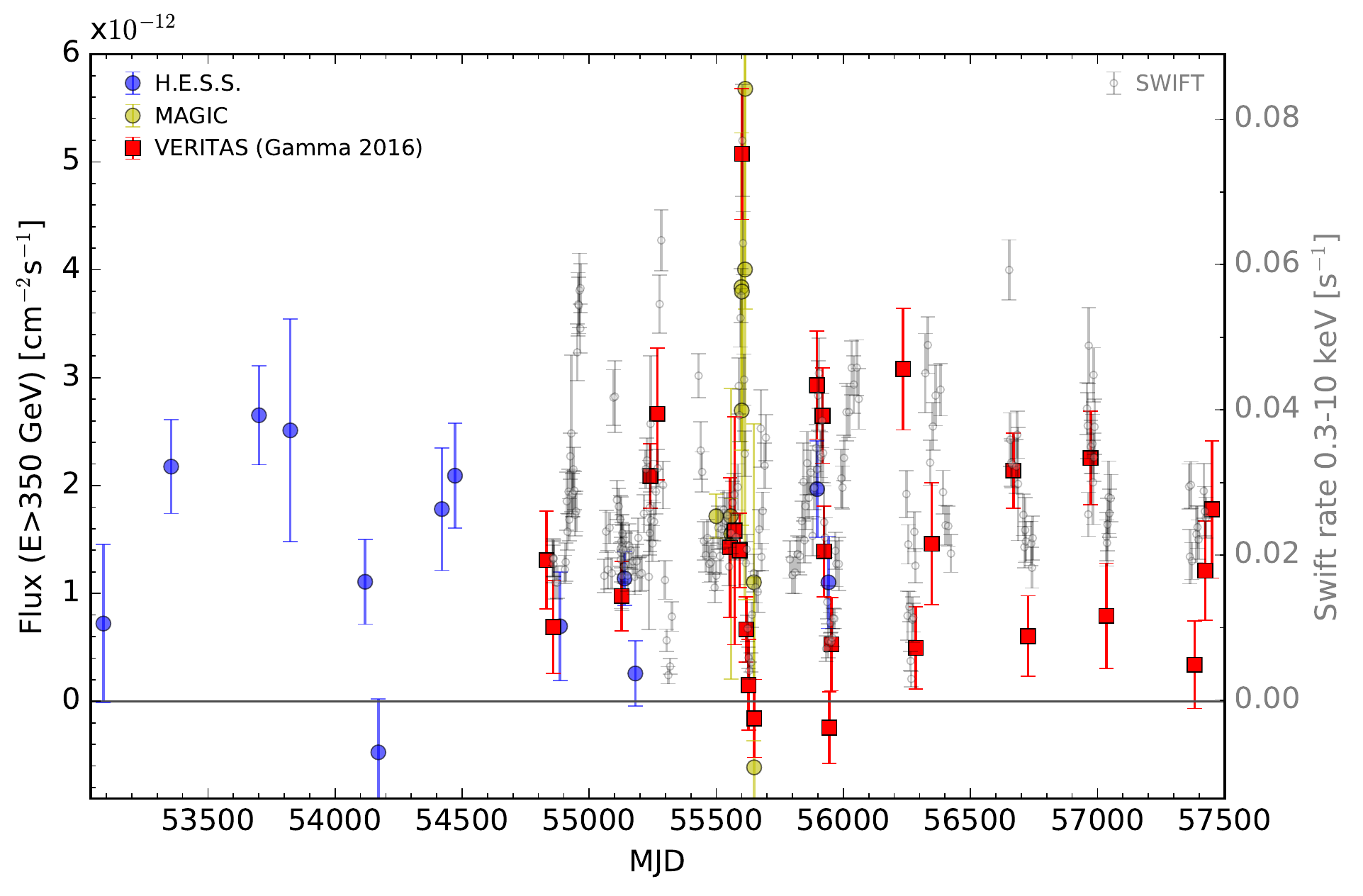}}
\caption{\label{fig:LT} Long-term gamma-ray curve of \hessj\ for energies $>$350\,\gev\ and X-ray light curve from {\em Swift} using data from 2004 to 2016. The last three red points show new data taken in 2015-16.}
\end{figure}

\section{INTRODUCTION}

The gamma-ray emitting binary \hessj\,(VER\,J0633$+$057) consists of a massive Be star 
(MWC\,148 = HD\,259440) and a compact object of unknown type (neutron or star black hole).
The system is located at a distance of 1.1-1.7\,kpc\,\cite{Aragona-2010}.
\hessj\ was discovered in the gamma-ray band  by H.E.S.S.\,\cite{Aharonian-2007}, serendipitously during observations of the Monoceros Loop supernova remnant as a point-like, variable 
source of very-high energy  ($>$100\,\gev) gamma rays \,\cite{Aharonian-2007,Acciaro-2009}. 
Long-term X-ray observations using the {\em Swift} X-ray Telescope (XRT) revealed periodic flux modulation. 
This points towards the binary nature of the system\,\cite{Bongiorno-2011} -- \hessj\ is considered one of the High Mass X-ray Binary systems.

\section{LONG-TERM OBSERVATIONS WITH VERITAS  AND {\em SWIFT}-XRT}

VERITAS is a gamma-ray observatory sensitive to photons in the energy range from 85\,\gev\ to $>$30\,\tev. It is located at the Fred Lawrence Whipple Observatory in southern Arizona (1300 meters above sea level, $31^{\circ}40'$N $110^{\circ}57'$W) 
and consists of an array of four imaging atmospheric-Cherenkov telescopes (details of the instrument and its performance were described previously\,\cite{Staszak-2015, Park-2015, VTS-specs}). 

The data used for these proceedings was recorded by VERITAS over a total of 208 hours between December 2006  and March 2016. The instrument went through several important changes during this period of time. The early data were taken during the construction phase of VERITAS with three telescopes. The array was completed in September of 2007  with the installation of the fourth telescope. In September of 2009  the original prototype telescope was relocated, leading to an improved sensitivity. In summer 2012 the cameras were upgraded, enabling the array to operate at a lower energy threshold. 

\begin{wrapfigure}{r}[0cm]{0.5\textwidth}
\captionsetup{width=0.47\textwidth}
\vspace*{-3ex}
\centerline{\includegraphics[trim={0 1.8ex 0 0},clip,width=0.5\textwidth]{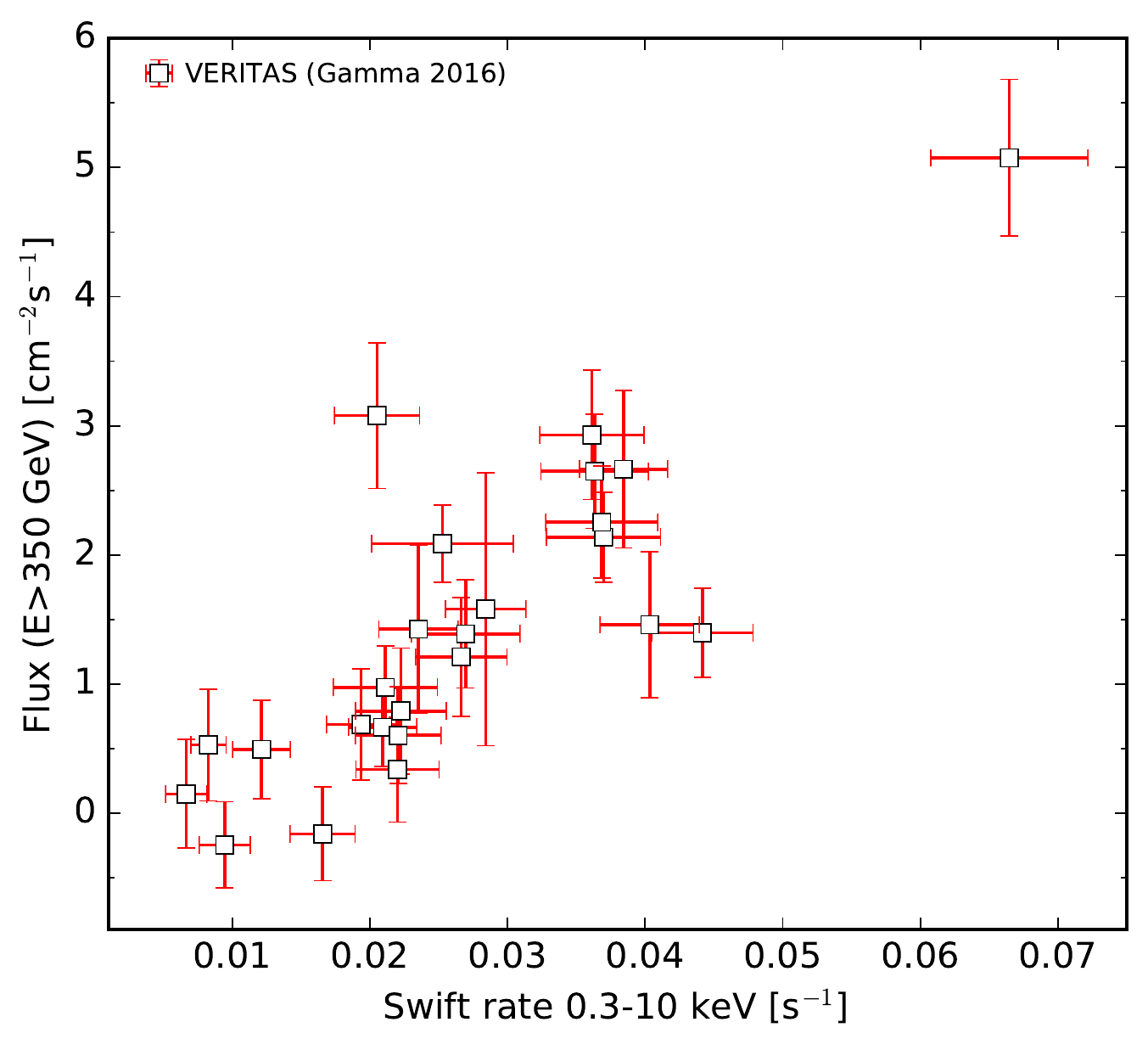}}
\caption{Gamma-ray fluxes ($>$350\,\gev) vs X-ray (0.3-10\,\kev) fluxes for contemporaneous observations. \label{fig:correl}}
\end{wrapfigure}
The {\em Swift}-XRT data cover the 0.3-10\,\kev\ band\,\cite{Burrows-2005}. {\em Swift}-XRT monitored \hessj\  from January 2009  to March 2016 with typical durations of $\sim$4-5\,ksec. While \hessj\  is clearly detected by {\em Swift}-XRT  it is the only gamma-ray binary which has not been observed at \mev--\hair\gev\ energies with  {\em Fermi} 
LAT\,\cite{Caliandro-2013}. 

In this note new measurements of the gamma-ray binary with the VERITAS telescopes obtained during observations in season 2015-2016 are presented. The VERITAS gamma-ray curve for energies above 350\,\gev\ is shown in Fig.\,\ref{fig:LT} together with  previous measurements by H.E.S.S.\,\cite{Aliu-2014} and MAGIC\,\cite{Aleksic-2012}. The VERITAS results presented here have been obtained with updated analysis algorithms compared to previous results\,\cite{Aliu-2014}. The most important change is the application of a boosted-decision tree based gamma-hadron separation algorithm to the data, leading to a sensitivity improvement of roughly 15-20\%\,\cite{Park-2015}. The total detection signi\-ficance of the described  \hessj\ data is $21.3\,\sigma$ (using the for\-ma\-lism from Li and Ma\,\cite{Li-Ma-1983}, eq 17).

To study the correlation between gamma-ray and X-ray fluxes the X-ray data were selected within a $\pm$2.5 day interval around the VERITAS observing dates. The result is shown in Fig.\,\ref{fig:correl} 
for 25  contemporaneous observations. 

\section{PHASE-FOLDED {\em SWIFT}-XRT AND GAMMA-RAY OBSERVATIONS}

\begin{figure}[htb]
\includegraphics[width=0.47\textwidth,height=6.9cm]{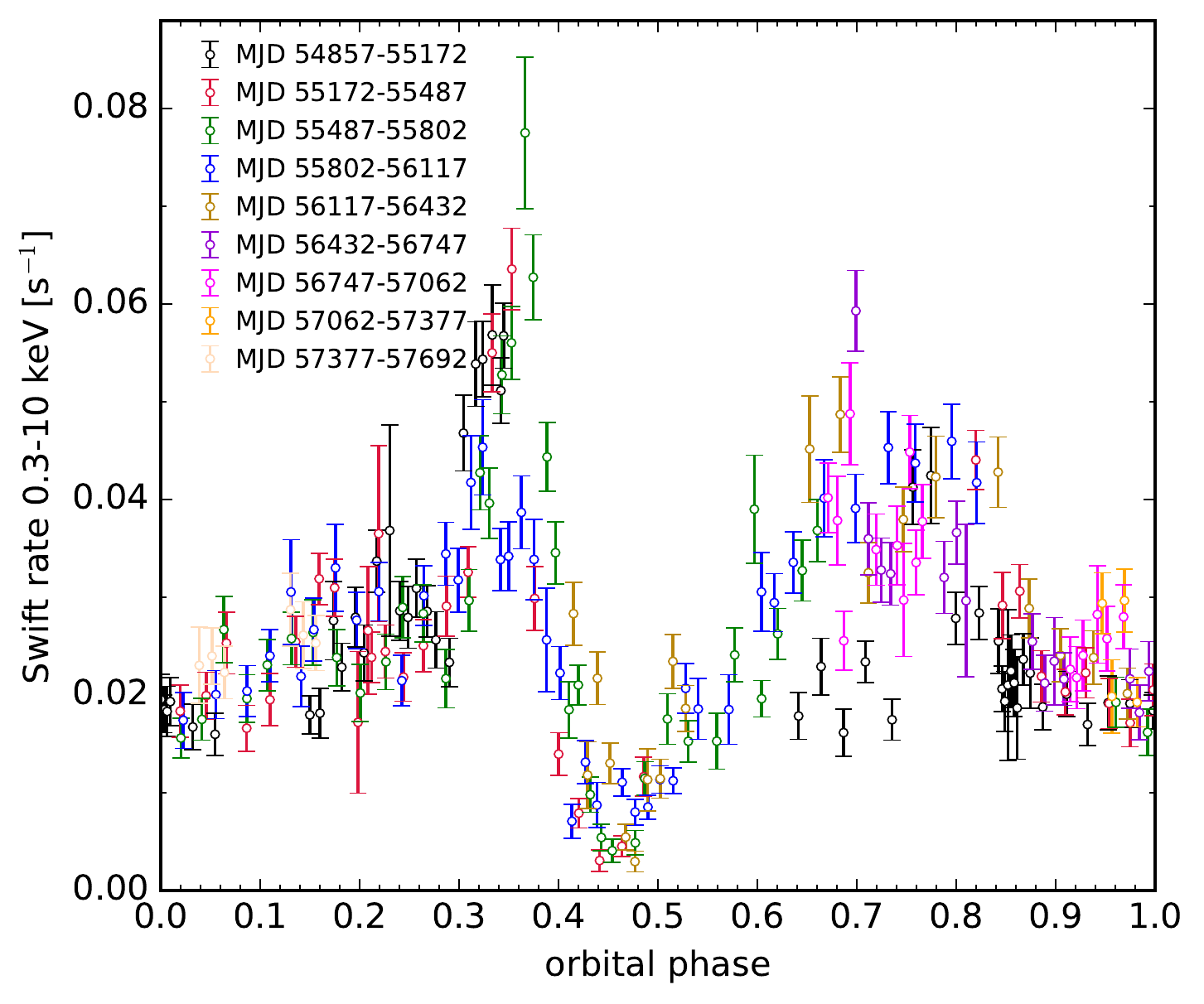}
\put(-22,160){a)}
\includegraphics[width=0.53\textwidth]{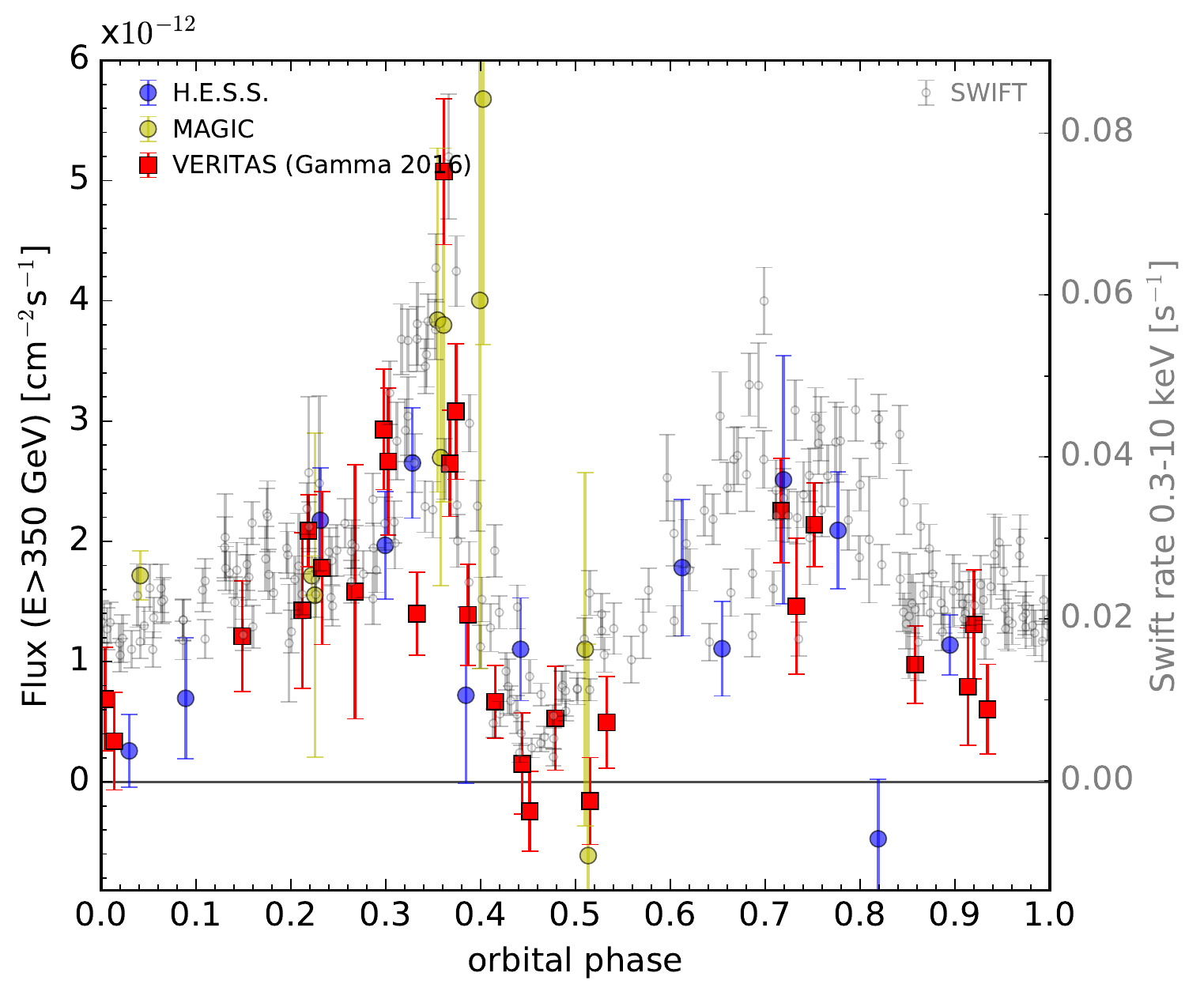}
\put(-48,160){b)}
\caption{Phase-folded X-ray and gamma-ray data using an orbital period of 315 days. a) {\em Swift}-XRT X-ray light curve in different orbital periods.
 b) Gamma-ray light curve for energies $>$350\,\gev\ overlaid with {\em Swift} data. 
\label{fig:phase}}
\end{figure}

Figure\,\ref{fig:phase} shows the gamma- and X-ray data folded with a period of 315 days. Both the X-ray data from {\em Swift} and the gamma-ray observations by VERITAS, H.E.S.S. and MAGIC reveal a pattern of variability with two prominent peaks around phases 0.35 and 0.75. A significant gamma-ray emission above 350\,\gev\ is visible. 

The X-ray rate versus time shows non-sinusoidal variations, thus the autocorrelation function 
(Z-DCF)\,\cite{ref:ZDCF1} and phase dispersion minimisation method\,\cite{ref:Stellingwerf1} were applied. Results are consistent with an orbital period of 315 days. The long orbital period of the binary of $315{\pm}5$ days\,\cite{Aliu-2014} makes it difficult to perform deep and continuous observations of the system during all orbital phases. It was
suggested that the orbit is eccentric ($e{\approx}0.83$) using radial velocity measurements of 
MWC\,148\,\cite{Casares-2012}. This puts the largest maximum of the gamma-ray emission at approximately phase 0.3 after the periastron -- clearly seen in Fig.\,\ref{fig:phase}. This pattern is similar to the binary LS\,I$+$61 303 (see e.g.\,\cite{Acciaro-2008}).

The variability pattern of the phase-folded gamma-ray light curve follows the X-ray light curve, as apparent 
in Fig.\,\ref{fig:phase}b). The first maximum at phases 0.2-0.4 is brighter than the emission at phases 0.6-0.8, but like the X-ray measurement exhibits orbit-to-orbit flux variability at similar phases, shown in Fig.\,\ref{fig:phase}a). The binary is also clearly detected in the phase range 0.8-0.2 (see  
Table\,\ref{table:results}). 

\section{PHASE-DEPENDENT SPECTRAL ANALYSIS}

\begin{figure}[htb]
\includegraphics[width=0.5\textwidth]{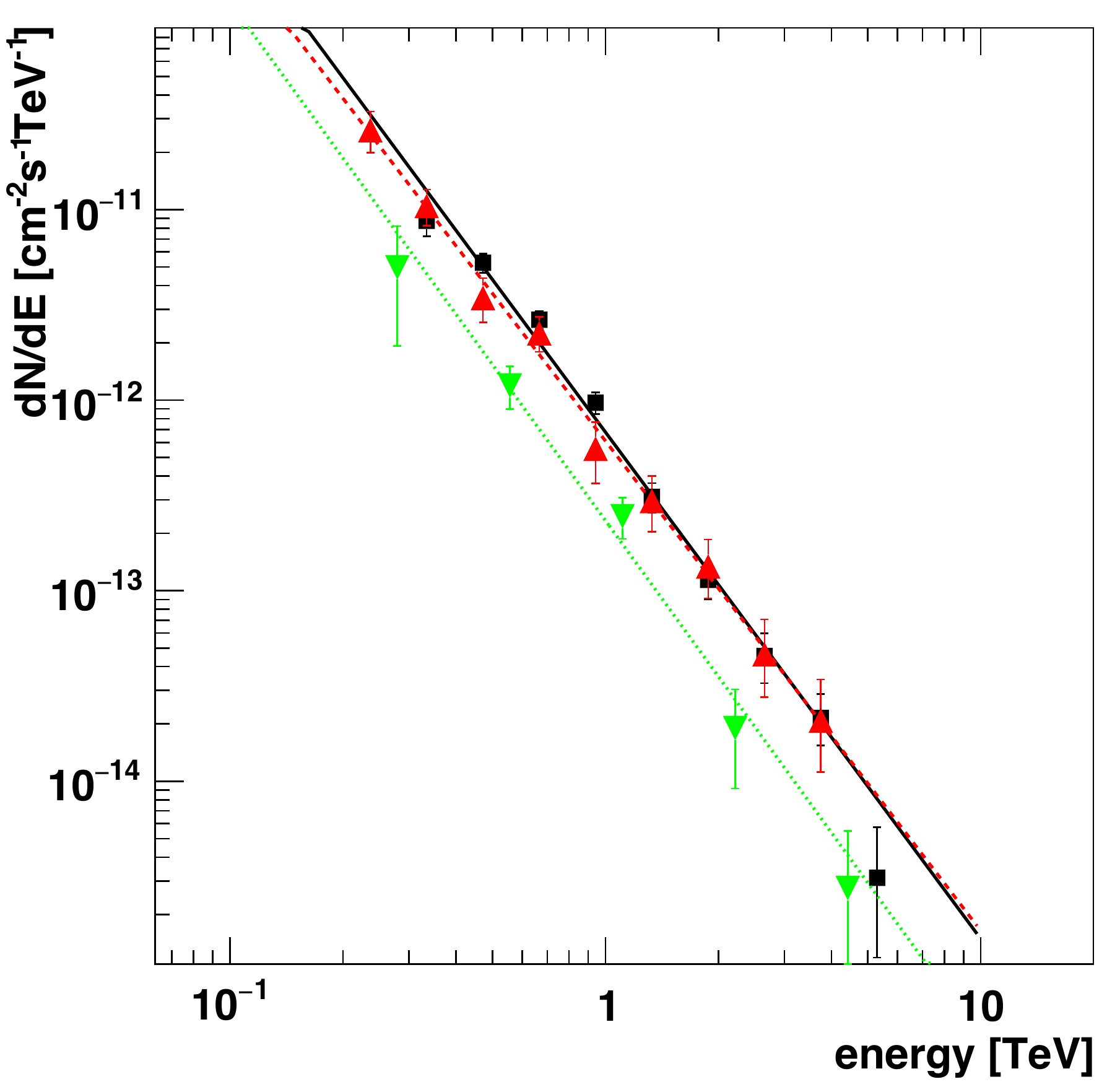}
  \caption{Differential energy spectra for gamma-ray photons in the orbital phase ranges of the maxima 0.2-0.4 (black) and 0.6-0.8 (red), and the elevated state 0.8-0.2 (green) -- see text for details.
  \label{fig:espec}}
\end{figure}
 
The differential energy spectra in gamma rays during the two maxima (phase ranges 0.2-0.4 and 0.6-0.8 respectively), and also during the elevated state  (phase range 0.8-0.2, as seen in Fig.\,\ref{fig:phase}b), can be described by power-law distributions. They are  shown in Fig.\,\ref{fig:espec}. The results of the spectral fits are consistent with each other (see results in Table\,\ref{table:results}), which points towards similar physical conditions at the gamma-ray emission sites during the high states before and after the apastron. 

\begin{table}[ht]
\caption{
\label{table:results}
Results of the spectral analysis results for energies $>$350\,\gev\ for the phase-folded \mbox{VERITAS} measurements in four phase ranges. 
A power-law distribution $dN/dE = {\it \Phi}_0 \cdot E^{-\gamma}$ of the data is assumed for the spectral fits,
see also Fig.\,\ref{fig:espec}.
All quoted errors are $1\sigma$ statistical errors only.
The flux normalisation constant  $\Phi_0$ is in units of $10^{-13}$ cm$^{-2}$ s$^{-1}$ \tev$^{-1}$.
}
\begin{tabular}{l|c|c|c|c|c}
\hline
{\bf Orbital phase} & {\bf 0.2-0.4}  & {\bf 0.4-0.6} & {\bf 0.6-0.8} &{\bf  0.8-0.2} \\
\hline
{\bf Observation time} (h) & 79 & 46 & 27 & 55 \\
{\bf Significance} ($\sigma$) & 19.2 & 2.6 & 11.5 & 6.8 \\
\hline 
{\bf $\Phi_0$ at 1\,\tev} & $6.8\pm0.4$ & - & $6.1\pm0.7$ & $2.3\pm0.4$ \\
{\bf Photon index} $\mathbf{\gamma}$ & $2.55\pm0.05$  & - & $2.57\pm0.12$  & $2.72\pm0.2$ \\
{\bf $\chi^2$/N} & 17.5/7 & - & 2.7/7 & 2.8/3 \\
\end{tabular}
\end{table}

\section{CONCLUSION}

The binary system \hessj\ was measured in the X-ray and gamma-ray band over ten years. Its light curve exhibits two maxima, with a minimum in the emission pattern close to the apastron. 
New and updated observations of the gamma-ray binary \hessj\ by VERITAS during the observing season 2015-2016 provide significantly improved  measurements of the complex situation in this gamma-ray binary over all phases of the binary orbit. VERITAS will continue to observe this source as part of its long-term observing plan.

\section{ACKNOWLEDGMENTS}
This research is supported by grants from the U.S. Department of Energy Office of Science, the U.S. National Science Foundation and the Smithsonian Institution, and by NSERC in Canada. We acknowledge the excellent work of the technical support staff at the Fred Lawrence Whipple Observatory and at the collaborating institutions in the construction and operation of the instrument. The VERITAS Collaboration is grateful to Trevor Weekes for his seminal contributions and leadership in the field of VHE gamma-ray astrophysics, which made this study possible.

This work made use of data supplied by the UK Swift Science Data Centre at the University of Leicester.


\nocite{*}
\bibliographystyle{aipnum-cp}%

\end{document}